%% file: main.tex
\documentclass[letterpaper]{article} 
\usepackage[]{aaai25}  
\usepackage{times}  
\usepackage{helvet}  
\usepackage{courier}  
\usepackage[hyphens]{url}  
\usepackage{graphicx} 
\usepackage{natbib}  
\usepackage{caption} 
\usepackage{amsmath}
\usepackage{svg}
\usepackage{multirow}
\usepackage{algorithm}
\usepackage{algpseudocode}
\usepackage{booktabs}
\usepackage{newfloat}
\usepackage{listings}
\usepackage{color}

\usepackage{enumitem}

\usepackage{algorithm}
\usepackage{algpseudocode}
\usepackage{float}
\usepackage{longtable}
\lstdefinelanguage{PDDL}{
  basicstyle=\ttfamily\small,
  keywords={:action, :parameters, :precondition, :effect, and, when, forall, assign, increase},
  keywordstyle=\color{blue}\bfseries,
  identifierstyle=\color{black}, 
  sensitive=true,
  comment=[l]{;},
  commentstyle=\color{gray}\itshape,
  morestring=[b]",
  stringstyle=\color{orange}
}

\usepackage{makecell}
\usepackage{longtable}
\usepackage{graphicx}
\definecolor{brown}{rgb}{0.6, 0.4, 0.2}
\usepackage{subcaption}

\title{Technology-assisted Personalized Yoga for Better Health - Challenges and Outlook}
\author {
    Vivek Kumar$^{1}$, 
    Himanshu Sahu$^{1}$,  
    Hari Prabhat Gupta$^{1}$,
    Biplav Srivastava$^2$ 
}
\affiliations {
    $^{1}$ IIT (BHU), Varanasi, India \\
    $^2$ University of South Carolina, Columbia, USA
}
\begin{document}
\nocopyright
\maketitle

\begin{abstract}
Yoga is a discipline of physical postures, breathing techniques, and meditative practices rooted in ancient Indian traditions, now embraced worldwide for promoting overall well-being and inner balance. The practices are a large set of {\em items}, our term for executable actions like physical poses or breath exercises, to offer for a person’s well-being. 
However, to get benefits of Yoga tailored to a person's unique needs, a person needs to (a) discover their subset from the large and seemingly complex set with inter-dependencies, (b) continue to follow them with interest adjusted to their changing abilities and near-term objectives, and (c) as appropriate, adapt to alternative items based on changing environment and the person's health conditions.
In this vision paper, we describe the challenges for the Yoga personalization problem. Next, we sketch a preliminary approach and use the experience to provide an outlook on solving the challenging problem using existing and novel techniques from a multidisciplinary computing perspective.  To the best of our knowledge, this is the first paper that comprehensively examines decision support issues around Yoga personalization, from pose sensing to recommendation of corrections for a complete regimen, and illustrates with a case study of Surya Namaskar - a set of 12 choreographed poses.
\end{abstract}

\input{content/introduction}
\input{content/background}

\input{content/challenges}
\input{content/solution}

\input{content/outlook}

\input{content/conclusion}

\section{Acknowledgments}
We thank Mr. Kush Pandey, Yoga Instructor at the Student Activity Center (SAC) of IIT(BHU), for sharing his expertise on Yoga and allowing us to measure the performance of volunteers. We thank Abhishek Verma, Gautam Bhagat, and Tanima Dutta for their help in data collection and analysis, and Shaheena and Vineela for preliminary implementation of REST APIs for sensory content management. We acknowledge funding of the VAIBHAV program for their partial support to Biplav Srivastava (visiting VAIBHAV fellow) and Hari Prabhat Gupta (hosting institute).

\bibliography{references}

\appendix
\input{content/appendix}
\end{document}

%% file: content/introduction.tex
\section{Introduction}  
Yoga, a traditional practice from India for holistic health~\cite{GARFINKEL2000125,griffith2013rig,ayush-yoga-info}, has been growing in popularity around the world. On 11 December 2014, the United Nations, through resolution 69/131~\cite{internationalyogaday}, declared June 21 as the International Day of Yoga, which has been celebrated annually since.
It has a large set of techniques to offer for a person’s holistic health across eight branches,
of which two - exercise (\={A}SANA) and breathing techniques (\={P}R\={A}\d{N}\={A}Y\={A}MA) - are integral elements of any regular Yoga regimen. The practices, which we call items, play a fundamental role in cultivating Yoga's holistic benefits, positively influencing both the physical and mental well-being of practitioners.

However, a common problem a person interested in Yoga faces is what items to undertake (practice) on a given day, with the time and their physical condition at that instance. Moreover, some may want changes over time to escape boredom from routine or find a better match to their custom physical and mental capabilities, while others may want to stick to a regimen. Today, information about Yoga is available in non-personalized, static media like documents or videos, and personalized guidance is only available at high cost with hard-to-find experts across the globe.
Furthermore, practicing yoga in an unsupervised manner can lead to mistakes which ultimately bring additional problems (aggravating pre-existing conditions) and potential health issues like sprains~\cite{yogasafe, yogasafe1, yogasafe2}. Existing resources, such as static media, documents, or videos, often do not monitor users to ensure they are performing the yoga poses correctly and provide feedback.

Our objective is to promote adherence to healthy living via Yoga in a privacy-preserving, personalized, and trusted manner.  Our research aims to develop a comprehensive system for promoting long-term healthy living through personalized Yoga practices, assisted by technology: Artificial Intelligence (AI), Ubiquitous Computing (UC), and Usage-Centered Human Computer Interfaces (HCI). The Artificial Intelligence (AI) problem involves recommending a subset of items to a user from the Yoga set. The Ubiquitous Computing (UC) problem is monitoring a user’s performance during Yoga via sensing. The Human Computer Interfaces (HCI) problem is designing effective interfaces that can deliver recommendations and monitoring solutions to a user at scale. Starting with Surya Namaskar - a set of 12 (\={a}sana) combined with breathing routines (\={p}r\={a}\d{n}\={a}y\={a}ma), repeated as sets – we want our approach to be scalable, and seek to promote a growing ecosystem of Yoga-based decision support tools, multidisciplinary research collaboration, and entrepreneurship to promote further innovations in this space.

There is a rich body of work on exercise monitoring, adaptation, and recommendation for individuals \cite{exercise-monitoring-survey,exercise-monitoring-survey-2} and specific demographics like seniors \cite{exercise-balancing-seniors}. There is also some work on machine learning methods for Yoga pose detection \cite{yoga-pose-detection,yogahelp-hari}. However, we are not aware of any work on Yoga personalization that comprehensively addresses the sensing of poses and recommendation of corrections for a large regimen of items, like the SN with 12 choreographed poses~\cite{sn-background}, to improve decision support for a practitioner.

%% file: content/background.tex
\section{Background and Related Work}
In this section, we provide preliminaries on Yoga, recommendations, and sensing methods that are needed to discuss the subject matter in depth. 

\begin{figure}[htbp]
    \centering
    \includegraphics[width=0.5\textwidth]{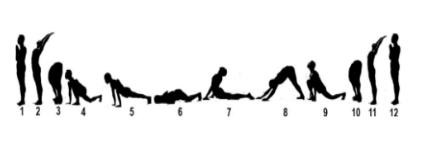}
    \caption{ Pictorial representation of the flow of postures of SN as per \textbf{Sivananda Yoga Vedanta Centre tradition}, which is practiced at IIT-BHU (Image source:~\citet{sn-background}). The pose names are: 1. Pr\={a}nam\={a}sana, 2. Hasta utth\={a}n\={a}sana, 3. P\={a}da hast\={a}sana, 4. A\'{s}vasanc\={a}lan\={a}sana, 5. Phalak\={a}sana or Caturanga da\d{n}\d{d}\={a}sana, 6. Aṣṭ\={a}\d{n}ga namask\={a}ra, 7. Bhujang\={a}sana, 8. Adho mukha \'s v\={a}n\={a}sana or Parvat\={a}sana, 9. A\'{s}vasanc\={a}lan\={a}sana, 10. P\={a}da hast\={a}sana, 11. Hasta utth\={a}n\={a}sana, 12. Pr\={a}nam\={a}sana.}
    \label{fig:sn-sivananda}
\end{figure}

\begin{figure}[htbp] 
    \centering
    \includegraphics[width=0.5\textwidth]{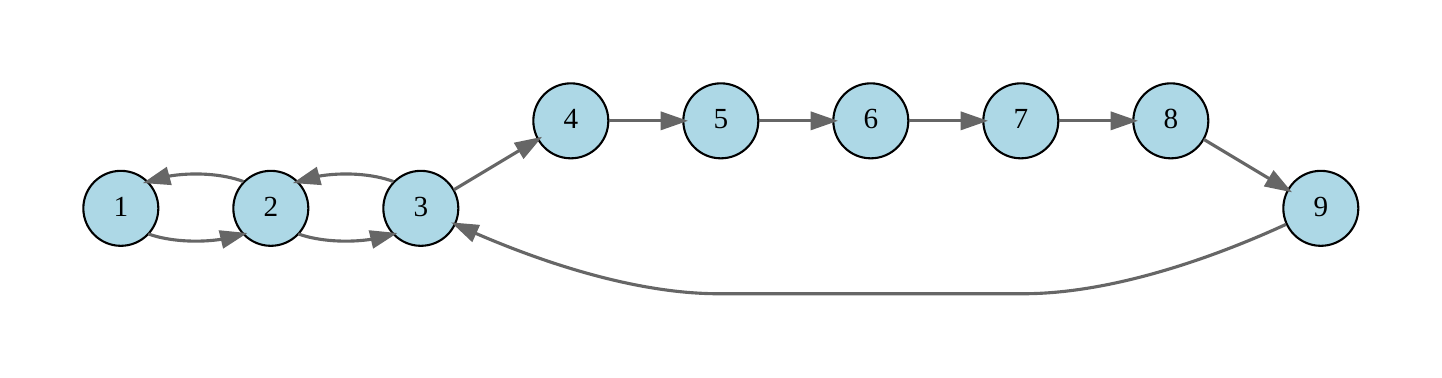} 
    \caption{Schematic representation of SN postures (Sivananda Yoga Vedanta Centre tradition).
}
    \label{fig:sn-state-transition-siva}
\end{figure}

\subsection{Yoga and Surya Namaskar}

Yoga has a large set of techniques to offer for a person’s holistic health, divided into eight branches: 
Y\={A}MA -- restraints, NIY\={A}MA -- observances, \={A}SANA -- posture or exercise, PR\={A}\d{N}\={A}Y\={A}MA -- breathing techniques, PRATY\={A}H\={A}RA -- sense withdrawal, DH\={A}RA\d{N}\={A} -- concentration, DHY\={A}NA -- meditation, SAM\={A}DHI -- enlightenment. We will focus on two of them - exercise (\={A}SANA) and breathing techniques (PR\={A}\d{N}\={A}Y\={A}MA) - and refer to them as items belonging to the Yoga set. These two integral elements of traditional Yoga activities play a fundamental role in cultivating its holistic benefits, positively influencing both the physical and mental well-being of practitioners.

Among Yoga practices, SN is considered a comprehensive set of choreographed poses that work on all parts of the body. \cite{sn-background} provides a history and survey of SN detailing the exercises (\={a}sanas) and associated information (order, breath control, and optional chanting) in the context of common SN choreographies from leading Yoga schools.
Figure~\ref{fig:sn-sivananda} depicts the authoritative Surya Namaskar sequence established by the Sivananda Yoga Vedanta Centre, serving as the foundation for our study~\cite{sn-background}. Diverse adaptations and deviations are presented in Figure~\ref{fig:sn-variations}.

\subsection{Recommendation and AI Techniques}

Recommendation is a subfield of AI that studies methods and algorithms for identifying and presenting items most likely to match a user’s preferences, needs, or context~\cite{isinkaye2015recommendation}. Most traditional recommender systems are built around a single user: they learn what a user prefers, predict what they might like, and present suggestions tailored just for them. These systems work remarkably well in contexts where a person browses alone—shopping online, watching videos, or reading articles \cite{li2023-recentdevelopments-recommendersystems}. Recent research in the field has focused on more complex situations where a {\em set} of items is recommended to a person, or to a {\em set of} users, or both. For example, in \cite{ultra-team-group-reco}, the authors consider recommending a funding opportunity to a group of people, and in \cite{mealreco-nagpal2024novelapproachbalanceconvenience}, a set of food items is recommended to a user.

In the context of Yoga, it can be viewed as a group recommendation problem where a set of items is to be recommended to a user subject to the latter's constraints and preferences.
\begin{figure*}[htbp]
    \centering
    \begin{subfigure}[t]{0.48\linewidth}
    \centering
    \includegraphics[width=0.6\linewidth]{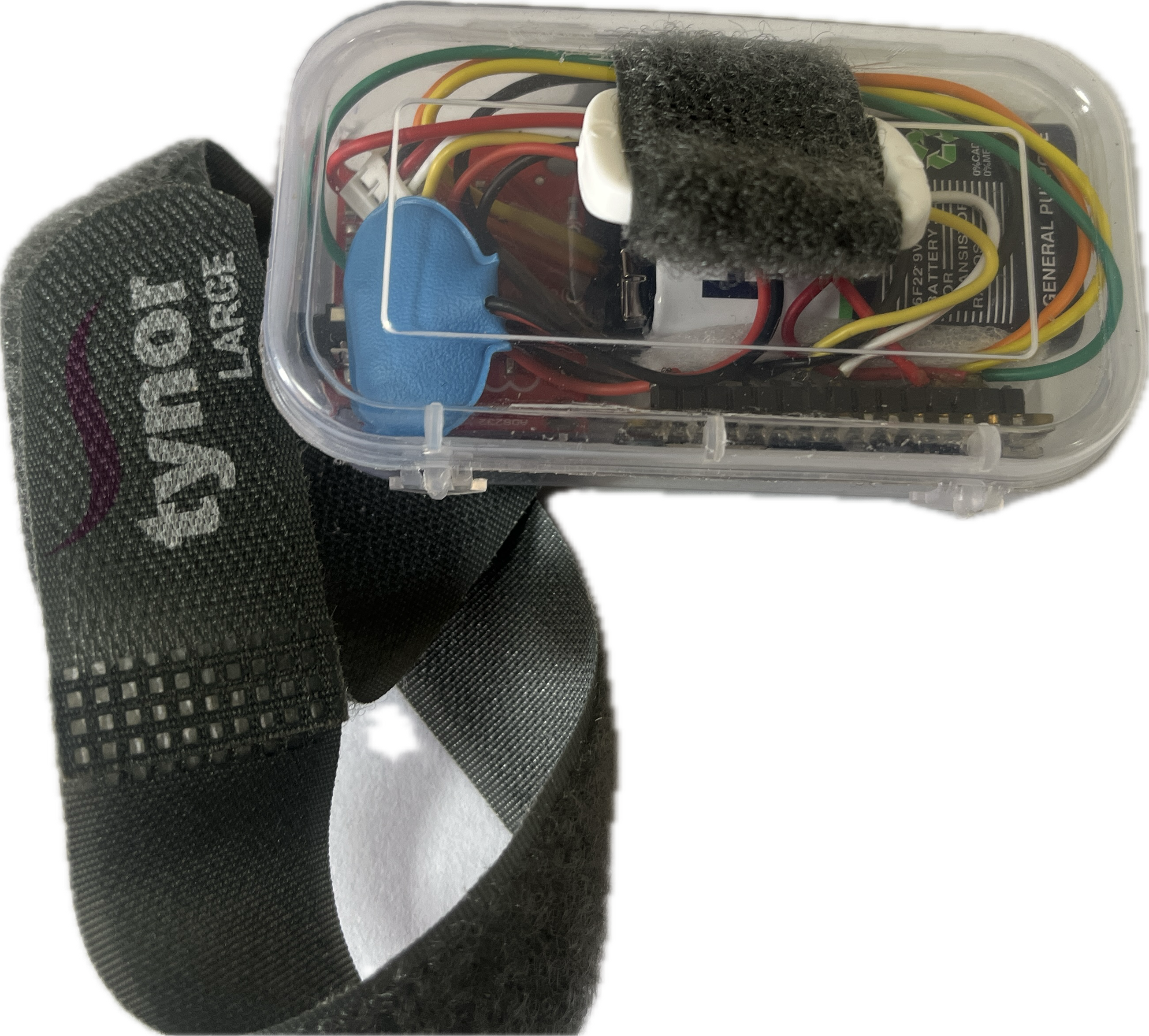} 
    \subcaption{Wearable Custom Sensor Node}
    \end{subfigure}
    \begin{subfigure}[t]{0.48\linewidth}
    \centering
     \includegraphics[width=0.8\linewidth]{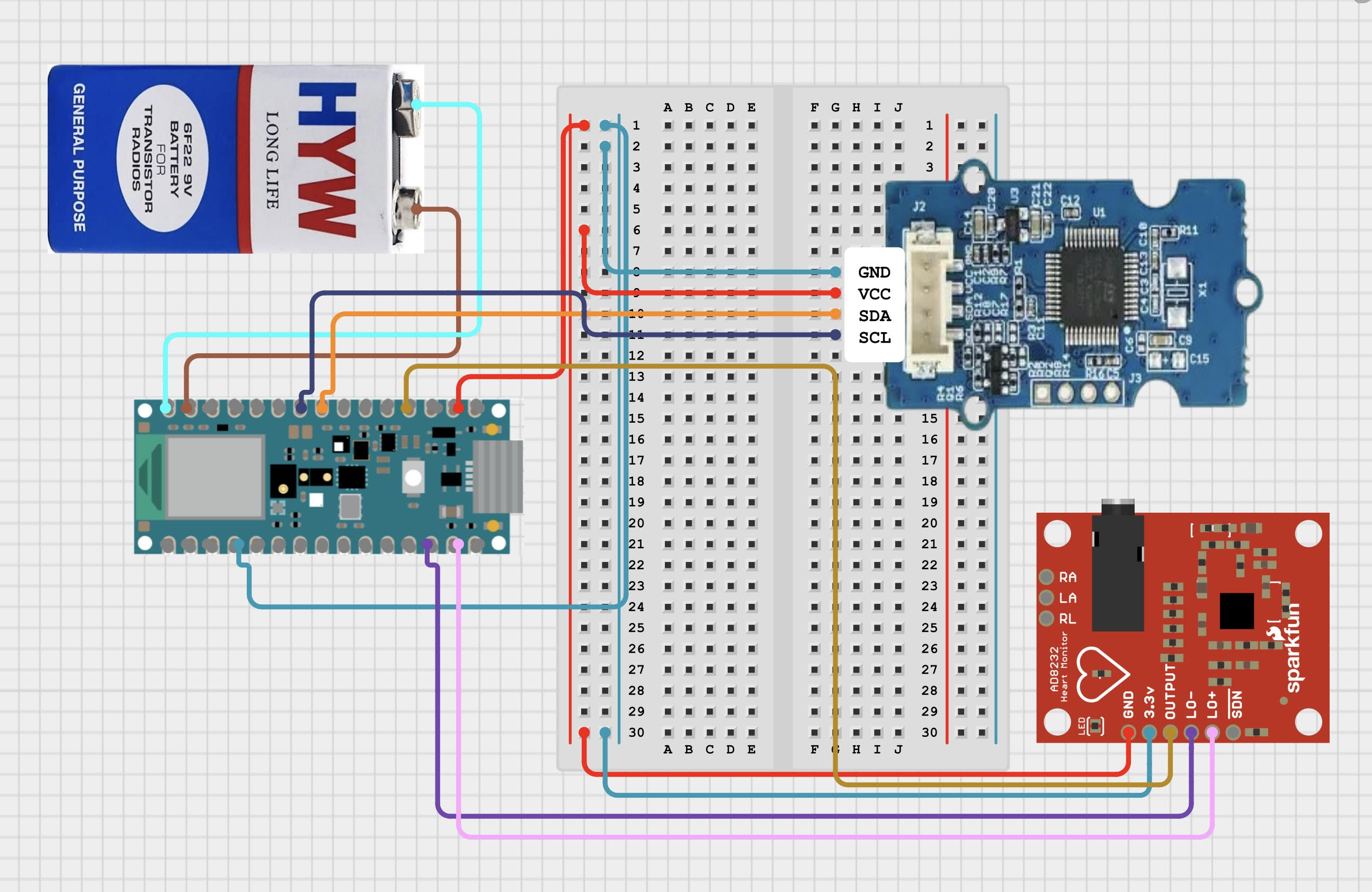}  
     \subcaption{Schematic circuit diagram}
    \end{subfigure}
    \caption{Custom wearable sensor node for yoga data collection. It integrates an Arduino Nano BLE module with a built-in IMU sensor, a heart-rate sensor, and a 9-volt battery for power. Designed for safe use, the device is worn on the hand just above the wrist during practice.}
    \label{fig:customnode}
\end{figure*}

\subsection{Sensing and Pervasive Computing}
Sensing and Pervasive Computing refers to a technology system where tiny sensors are used to collect information from the surroundings, and smart devices automatically use this information to help people in their daily lives, often without them even noticing. For example, a fitness watch that tracks your steps and sleep, or smart lights that turn on when you enter a room, are both examples of sensing and pervasive computing. Such devices are ubiquitous in nature and cause no resistance during daily activities. Based on the availability and nature of sensing, these devices can be either custom or off-the-shelf. Custom devices are application-specific, containing only the required components. In contrast, off-the-shelf devices include the most common components. The advantages of custom devices are that they are cost-effective, energy-efficient, and easy to customize. However, due to their compact size, the user experience may not be as appealing compared to off-the-shelf devices.

\subsubsection{Custom Board} The Arduino Nano BLE is a compact microcontroller board integrating an ARM Cortex-M4 processor and an in-built IMU sensor, making it suitable for motion and orientation tracking. Its low power consumption and small form factor enable unobtrusive wearable sensor node deployment. Bluetooth Low Energy (BLE) provides wireless data transfer with minimal energy usage, ideal for battery-powered wearables~\cite{9283231} . It supports short-range, low-latency communication between the sensor node and a paired device for real-time yoga pose monitoring. 
A custom Bluetooth Low Energy (BLE) sensor, designed using an Arduino Nano BLE module~\cite{yogahelp-hari}, can be worn and secured to the wrist (as shown in the Figure~\ref{fig:customnode}), enabling the capture of Inertial Measurement Unit (IMU) readings during movements. 

The Arduino Nano BLE integrates a 9-axis IMU (LSM9DS1) combining a 3-axis accelerometer, 3-axis gyroscope, and 3-axis magnetometer in a single package. This sensor enables precise motion tracking, orientation estimation, and angular velocity measurement. In yoga applications, the IMU can capture subtle postural shifts, detect alignment deviations, and log dynamic transitions between poses for real-time feedback or post-session analysis.
One can augment initial sensing with common sensors like an accelerometer and a gyroscope by expanding  to a variety of upcoming sensors as well, like heart rate and bio-signal monitoring sensors, such as:
\begin{itemize}
    \item AD8232 ECG Sensor\footnote{\url{https://learn.sparkfun.com/tutorials/ad8232-heart-rate-monitor-hookup-guide/all}}: The AD8232 ECG sensor module is a low-power, single-lead heart rate monitor designed for measuring the heart's electrical activity. It amplifies and filters small biopotential signals, providing an analog output compatible with microcontrollers like Arduino. It is a valuable tool for DIY health projects, wearable devices, and educational purposes. 
    \item Grove Finger-clip Heart Rate Sensor\footnote{\url{https://wiki.seeedstudio.com/Grove-Finger-clip_Heart_Rate_Sensor/}}: It is a compact, wearable sensor that uses a PAH8001EI‑2G optical heart‑rate detection chip with an integrated green LED and DSP, leveraging photoplethysmography to measure blood‑flow variation. It offers ultra-low power consumption with sleep modes, an I²C interface, and an embedded STM32 microcontroller (with SWD interface) for on-chip processing and reprogramming. Ideal for deployment on the finger or wrist, this sensor includes a 3D‑printable shell and bandages for secure and accurate placement. 
\end{itemize}

\subsubsection{Off the Shelf Node} 
\begin{figure}[htbp]
    \centering
    \begin{subfigure}[t]{0.48\linewidth}
        \centering
        \includegraphics[width=\linewidth]{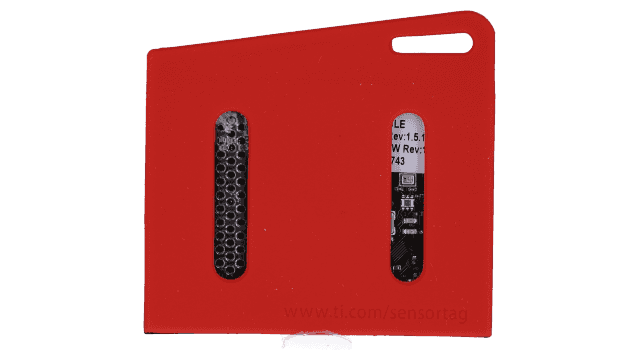}
        \caption{Outer view}
        \label{b}
    \end{subfigure}
    \hfill
    \begin{subfigure}[t]{0.48\linewidth}
        \centering
        \includegraphics[width=\linewidth]{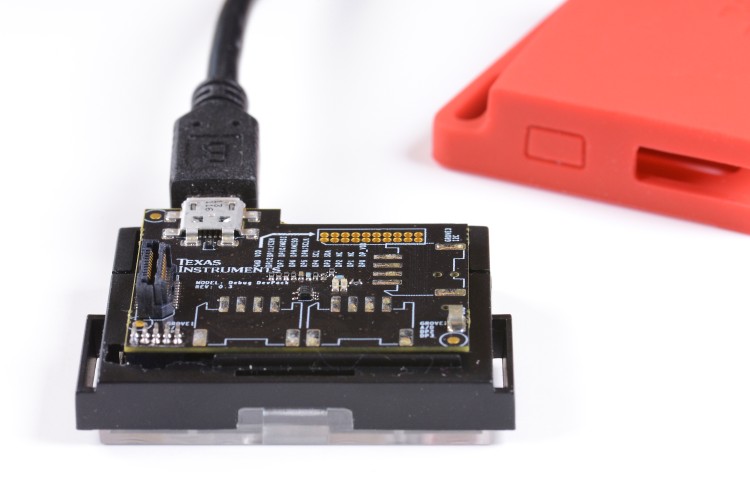}
        \caption{TI sensor tag board}
        \label{a}
    \end{subfigure}
    \caption{Texas Instruments CC2650STK SensorTag used for data collection, equipped with integrated motion, environmental, and ambient sensors for wireless monitoring of yoga postures.}
    \label{fig:tisensortag}
\end{figure}

The Texas Instrument Sensor Tag (CC2650STK)~\cite{tisensor} is a compact, battery‑powered IoT evaluation board based on the low‑power wireless MCU, featuring ten integrated sensors—including IMU, temperature, humidity, pressure, and ambient light—packaged in a small, red form factor. Figure~\ref{fig:tisensortag} shows the outer look of the sensor tag and the board on the right. It is an ultra-low power consumption device that can enable multi-year operation on a single coin-cell battery. This wide array of motion, orientation, and environmental sensors makes the Tag well-suited for tracking body posture and movement during yoga. Its Bluetooth LE connectivity enables real-time streaming of pose data to mobile or cloud apps for live feedback, pose correction, and performance analysis.

%% file: content/challenges.tex
\section{Challenges in the Yoga Context}
The desiderata for a system to provide personalized Yoga information are that it should be:
\begin{enumerate} 
    \item Accurate - identify the Yoga items correctly and unambiguously from sensed data
    \item Reliable - be able to handle users of diverse backgrounds (e.g., body types and conditions)  robustly
    \item Cost-effective - from installation to maintenance and total cost of ownership
    \item Non-invasive - not overwhelm the Yoga practitioner. At least,  the system should be non-obtrusive to the practitioner's activities.
    \item Power-efficient  -  the system should be long-running without energy recharge considerations, which is a major source of maintenance costs.
\end{enumerate}

There are many ways to achieve a particular goal for each above. For example, for accuracy, one can use a sensor with a high sampling rate. For low energy, one can use a BLE-based system that collects data from 4–5 sensors on a single laptop or tablet. But these ways may interfere with the set of overarching goals, \textit{e.g.}, need for high accuracy against low energy. Hence, there is a need for systematic research to meet the desiderata as much as possible.

\subsection{Data Challenges}

There are many data-related challenges around Yoga and SN. We cover a few of them here.
\begin{itemize}
    \item \textbf{Unfamiliar and obscure names of poses}: 
    Names of Yoga items have been derived from ancient Sanskrit texts, and they can be hard to pronounce. Furthermore, many items in different variants of Surya Namaskar referring to the same pose are named differently, which may cause confusion. For example, ‘PR\={A}NAM\={A}SANA’ is referred to as ‘SAMASTHIT\={I}’ in Variant-1, but executed differently; hence, it was recorded separately as Pose 18 in Table~\ref{tab:full-sn-siva}. Similarly, ‘HASTA UTT\={A}N\={A}SANA’ is called ‘T\={A}D\={A}SANA’ in Variant-1 but differs in execution and is labeled as Pose 17. This inconsistency in naming makes it challenging for learners, especially beginners, to understand or relate poses across different traditions.

    \item \textbf{Many many variants of SN and Yoga}: Pose details from multiple sources, including SN-base, Variant-1, Variant-2, and Variant-3, revealing significant variation in posture sequences, names, and execution styles. For instance, ‘Caturanga da\d{n}\d{d}\={a}sana’ appears in different forms across the base and variants—sometimes called ‘Phalak\={a}sana’ but performed differently in each variant. Such diversity, while showcasing richness in yogic traditions, also creates a fragmented understanding for practitioners trying to follow a standardized or unified practice.
    
    \item \textbf{Providing deeper domain (bodily) and semantic context}:  Each Yoga item impacts some regions of the human body.   For understand these deeper bodily and semantic relationships, one needs to m Yoga items to physical body parts and other physiological systems so that one can assess impact of SN on the practitioner. 
    One can also link each pose to specific Chakras, a Yogic concept that ties physical body parts to energy centers. For example, ‘P\={a}da hast\={a}sana’ is associated with the Swadhisthana Chakra (located in the lower abdomen), and ‘A\'{s}va Sanc\={a}lan\={a}sana’ with the Ajna Chakra (linked to the forehead or third eye area). This mapping helps in understanding the physiological and spiritual impact of each pose, enabling better awareness of how yoga influences different parts of the body.
\end{itemize}

\subsection{Sensing Challenges}

The challenges in sensing are about what to measure – physical exercise aspects (incline, speed), breath, environmental (moisture/sweat)?
How to measure?
What to use as a baseline – image/video?
What sensors? How to set up? What multi-sensory and node approach to follow?
How to store and retrieve consolidated data at scale collected? 
What data protocol to follow for user data? 
How to engage users in improving data practices and analysis?
Challenges in Yoga item detection are also many. We list some of them here. 
\begin{itemize}
    \item {\bf Multi modality}: Multimodal data collection requires multiple devices, and any fault or sensor error results in missing data points. Since the experimental data come from a live yoga session in an uncontrolled environment, collection is challenging. Table~\ref{Tab:datascale} presents daily collection statistics by modality and scale, highlighting increasing complexity with data volume.
    \item {\bf Data Labeling}: There are no specialized tools available for data labeling for multimodal data, so mostly manual labeling is used, which is time and effort-consuming. 
    \item {\bf Data Storage Retrieval}: The multimodal data for Y4all contains significant data points to call it big data, and storage and retrieval of such data requires specialized techniques for the same. 
    \item {\bf Synchronisation}: Multi-source as well as multimodal data synchronisation is a key challenge in processing the data. 
\end{itemize}

\subsection{Analytical Challenges}

The aim of analyzing collected data on a practitioner is to provide them with meaningful insights on how to improve their performance. For example, (1) the person may want to know how well their particular session was or (2) how to improve their practice for different poses based on their past performance (relative basis of comparison) or with respect to a Yoga expert (absolute basis of comparison). Furthermore, there are even more subtle challenges like (3) what to tell a person about their Yoga performance during a session (e.g., only errors or reinforce good poses) and (4) how to compute the information (analysis) needed for the insight from the collected data? 

There are further challenges at a much more detailed level. For example: (5) what  SN representation to use for data and resulting analysis, (6) how to detect pose (of Asana/ item) 
from image/ video or from sensor data or collectively, (7) how to nudge a person on correction while they are performing SN, (8) how to identify and recommend alternative  
poses (asana), and (9) what metrics to use to evaluate SN. As we see from the above discussion, Yoga provides us with a rich context to provide personalized information to improve a practitioner's performance.

%% file: content/solution.tex
\begin{figure*}[htbp]
    \centering
    \includegraphics[width=0.9\linewidth]{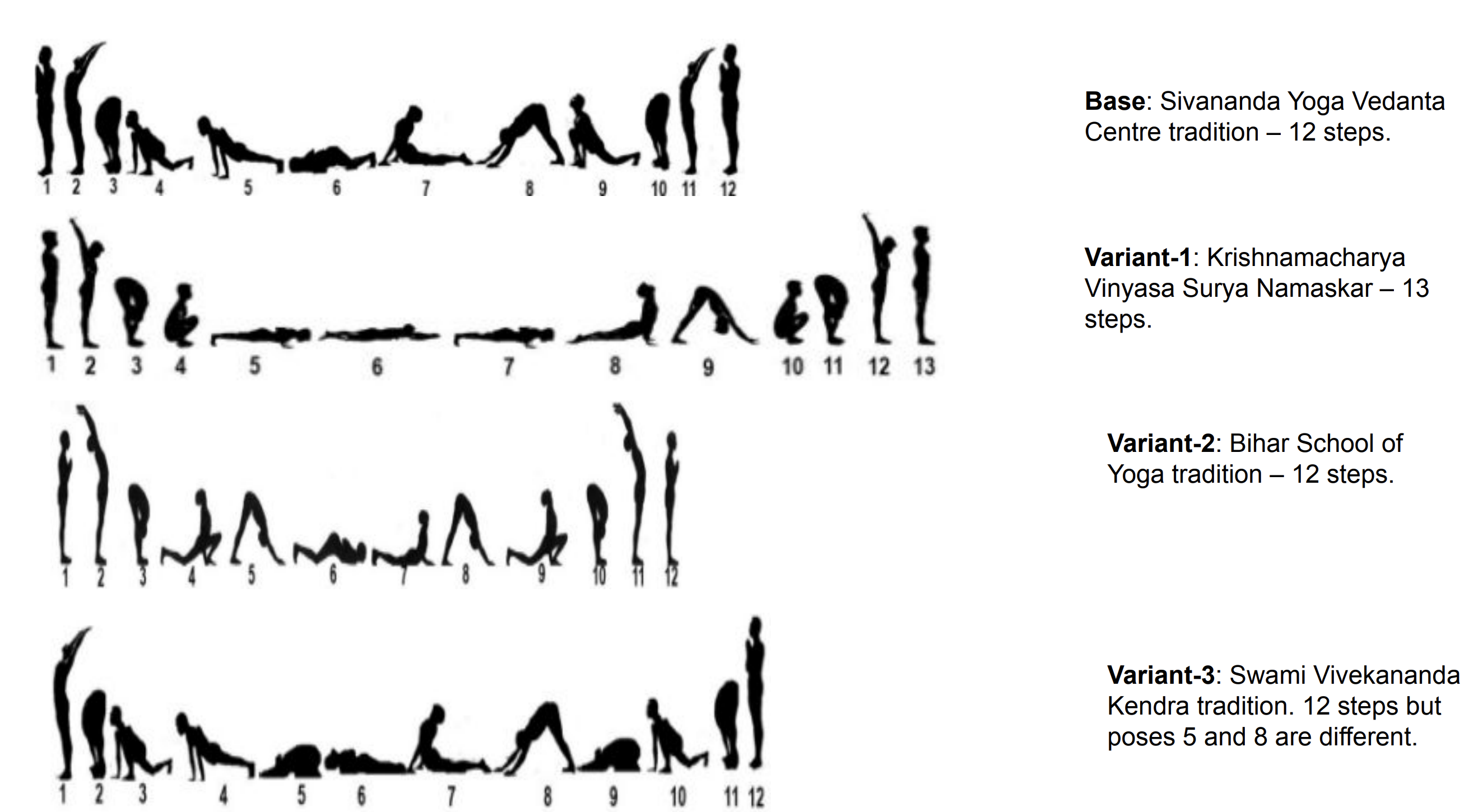} 
    \caption{Pictorial representation of SN variations (Image source:~\citet{sn-background}).}
    \label{fig:sn-variations}
\end{figure*}

\begin{figure*}[htbp]
    \centering
    \begin{subfigure}[b]{0.9\linewidth}
        \centering
        \includegraphics[width=\linewidth]{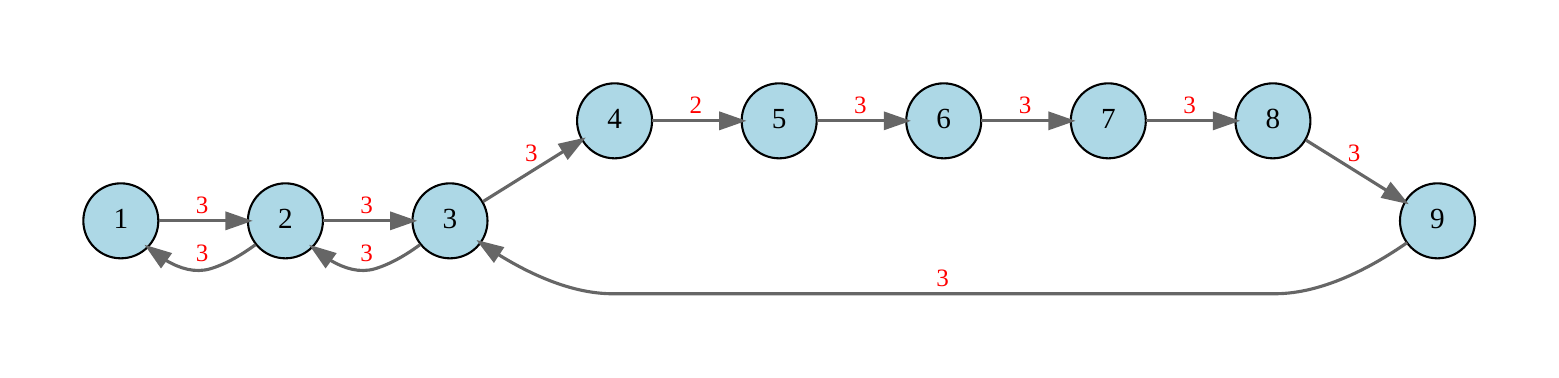}
        \caption{ Weighted DAG of 23rd July 2025, SN Pose, Set 1, Round-1. Total weight (time  in seconds) is  32.}
        \label{fig:subfig1}
    \end{subfigure}
    \hfill 
    \begin{subfigure}[b]{0.9\linewidth}
        \centering
        \includegraphics[width=\linewidth]{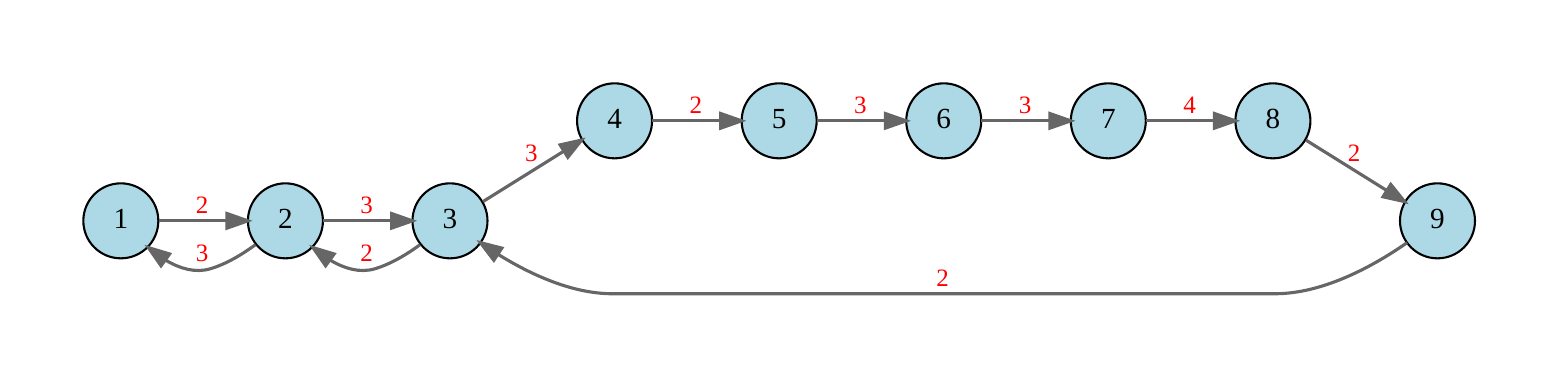}
        \caption{Weighted DAG of 23rd July 2025, SN Pose, Set 1, Round-2. Total weight (time  in seconds) is  29.}
        \label{fig:subfig2}
    \end{subfigure}
    \caption{Weighted Directed Acyclic Graphs (DAGs) (from videos) of SN Poses.}
    \label{fig:mergedfig-sn-session-analysis}
\end{figure*}
\section{A Preliminary Approach}

We now describe a preliminary approach that we implemented during a month at IIT-BHU at Varanasi, India, between July and August 2025. This experience will help us ground the discussion as we consider the future outlook in the next section.

\begin{figure}[htbp]
    \centering
    \includegraphics[width=0.8\linewidth]{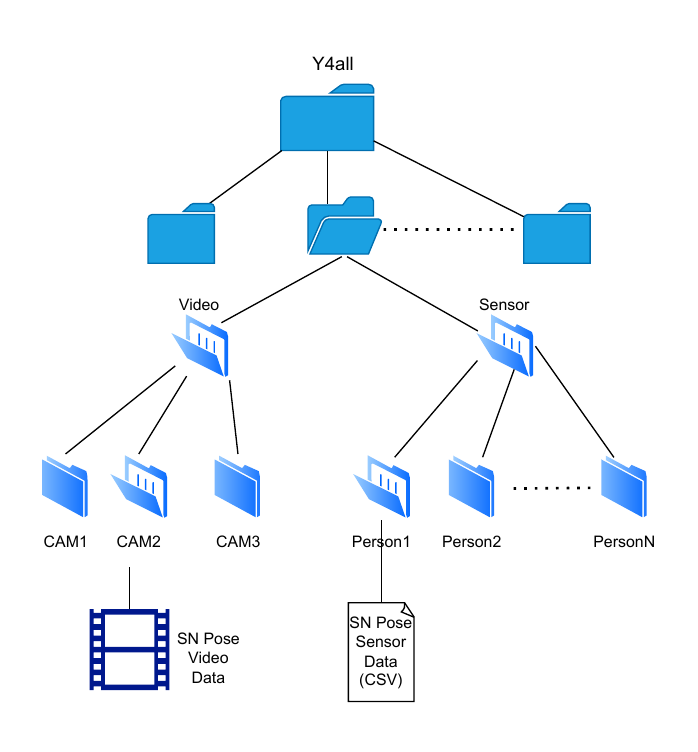}
    \caption{Directory Structure of the Dataset}
    \label{fig:filestore}
\end{figure}
\subsection{SN Data Consolidation}
We have created the SN items (pose) details by referring to and incorporating insights from the following two research papers:

1. {\em ``YogaHelp: Leveraging Motion Sensors for Learning Correct Execution of Yoga with Feedback" \cite{yogahelp-hari}} – This paper provided technical insight into pose recognition and correctness assessment in Sun Salutation using motion sensors. We referred to it particularly for identifying the correct sequence of poses, variations, and feedback mechanisms associated with posture correction.

2. {\em ``Insights on Surya Namaskar: From its Origin to Application Towards Health" \cite{sn-background}} – This review article was used for understanding the traditional background, different school-based variations (like BSY, Krishnamacharya, Sivananda), associated mantras, chakras, and detailed naming of poses.

Using these two resources, we documented the pose names, chanting mantras or beeja mantras, sources (e.g., base, variant traditions), associated chakras, and additional references. The aim was to provide a unified comparison across different SN traditions and highlight variations in pose execution, naming, and spiritual aspects.
These references were essential in accurately identifying and mapping the variations of postures, pose numbers in each tradition, and breath-mantra coordination. We also included annotations where the same pose had different names or execution styles across traditions (e.g., T\={a}\d{d}\={a}sana vs. Hasta utth\={a}n\={a}sana).
The list of SN variations is shown in Figure~\ref{fig:sn-variations} and details of the one we consider (Sivananda school) are shown in Table~\ref{tab:full-sn-siva}.
\subsection{Multi-modal Sensor Data Collection and Management}
\begin{table*}[!ht]
\centering
\footnotesize
\caption{\label{Tab:datascale} Modalities and scale of dataset collected per SN session of approximately 10 minutes.}
\begin{tabular}{|l|l|l|l|}
\hline
\textbf{S. No.} & \textbf{Modality} & \textbf{Data Scale} & \textbf{Comments} \\
\hline
\multirow{2}{*}{1.} & \multirow{2}{*}{Video Recording} 
&$\sim$1.8\,GB (= 3 $\times$ 0.6\,GB per camera) & 3 cameras (SN-only) \\
& & $\sim$12\,GB (= 3 $\times$ 4\,GB per camera) & 3 cameras (entire session) \\
\hline
2. & Custom Sensor & $\sim$1\,MB (20K+ IMU rows) & 1 sensor for SN duration \\
\hline
3. & Off-the-shelf Sensor & $\sim$1\,MB (20K+ IMU rows) & 2 sensors for SN duration \\
\hline
4. & Photos & $\sim$5\,MB & 0--6 images captured \\
\hline
\end{tabular}
\end{table*}

\subsubsection{Multi-modal Sensor Data Collection}

In our setup, we used a customized sensor device that includes Bluetooth Low Energy (BLE) and strapped it to the wrist while performing Surya Namaskar (SN) poses. This allowed us to capture IMU (Inertial Measurement Unit) readings during the movements. At the same time, a team member monitored the system and recorded all the data on a device in real time. Alongside the custom sensor, we also used an off-the-shelf sensor known as the TI Tag Sensor to measure IMU data. Although it's a commercial product, we used it in a way similar to our customized setup to maintain consistency. All measurements were taken during official yoga classes at IIT (BHU) SAC, 2nd Floor, under the guidance and supervision of a certified yoga instructor. The instructor led the sessions by guiding the SN poses step by step, and we followed accordingly. The entire session was video recorded, so we can accurately map each pose with the corresponding sensor data.

\subsubsection{Multi-modal Sensor Data Storage and Retrieval}

The dataset is organized using a hierarchical folder-based architecture (as shown in Figure~\ref{fig:filestore}) to manage multimodal sensor data efficiently. At the root level, the dataset is divided into three primary categories: video recordings, IMU sensor data, and pose estimation data in CSV format. The Video directory contains subfolders for each camera (CAM1, CAM2, and CAM3), where raw video files are stored. The IMU sensor data directory is structured by subject, with folders such as Person1, Person2, up to PersonN, each containing the respective individual's IMU sensor readings. In parallel, the SN Pose Sensor Data (CSV) directory provides synchronized pose information for each person in a structured tabular format. This modular design enables clear separation of modalities and supports efficient indexing, synchronization, and analysis of sensor data across subjects and sessions.

Data access is facilitated through a RESTful API, allowing programmatic retrieval of video, IMU, and pose data. The API endpoints are designed to support query parameters for specifying camera ID, person ID, session timestamps, and data modality. For instance, clients can request specific video files from a chosen camera, IMU readings for a particular individual, or pose data in CSV format filtered by session. This approach ensures scalable and secure access to large volumes of sensor data, supporting use cases such as real-time streaming, offline analysis, and machine learning pipeline integration. The REST API enhances the dataset's usability across distributed systems and enables seamless data ingestion for downstream processing tasks.\\

\subsection{Analysis}
The focus in the current approach was on assembling the data collection and analysis pipeline. Hence, the analysis task was scoped to construct a weighted state-transition diagram that could summarize an SN session. 

Figure~\ref{fig:mergedfig-sn-session-analysis} shows the result of such an analysis for two SN regimens. The weights represent the seconds spent when transitioning between the items (poses). Even the two regimens differ in their weights/ time taken. One may also compare these with the standard SN state transition (Figure~\ref{fig:sn-state-transition-siva}) to do a high-level comparison.

Note that the analysis itself was performed manually, which can be automated in the future. We also left  SN (Yoga) personalization as future work.

%% file: content/outlook.tex
\section{Future Outlook}

We now provide an outlook on how the preliminary implementation could be extended forward and the issues involved. We organize the discussion starting with analysis and then proceeding to sensing and computing platforms.

\subsection{Analysis and Personalized Recommendation}

Analysis is a key task of Yoga personalization, which includes the application of different AI and machine learning (ML) techniques over the collected user data as per the decision support requirements. The key challenge of data analysis is the ground truth estimation for which different approaches can be used: 
\begin{itemize}
    \item Manual Pose Marking: Video annotations to mark the representative pose of SN.  
    \item Sensor-Based Activity Recognition: The IMU sensor data can be used to establish ground truth based on the motion data. 
\end{itemize}
\begin{figure*}[htbp]
    \centering
    \begin{subfigure}[t]{0.48\linewidth}
        \centering
        \includegraphics[width=0.7\linewidth]{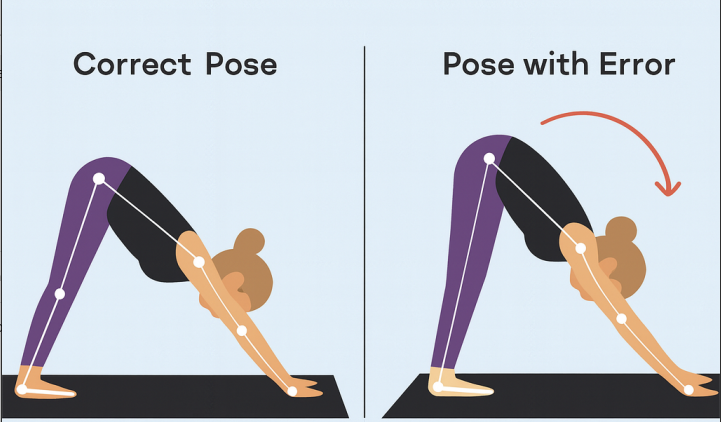}
        \caption{The user's (aggregated) previous performance (relative basis) or that of an expert (absolute basis) is shown on the left, while the current performance is given on the right. The comparison provides an error is the pose  (reference Pose 8 from SN - Table~\ref{tab:full-sn-siva}).}
        \label{fig:pose-error}
    \end{subfigure}
    \hfill
    \begin{subfigure}[t]{0.48\linewidth}
        \centering
        \includegraphics[width=0.6\linewidth]{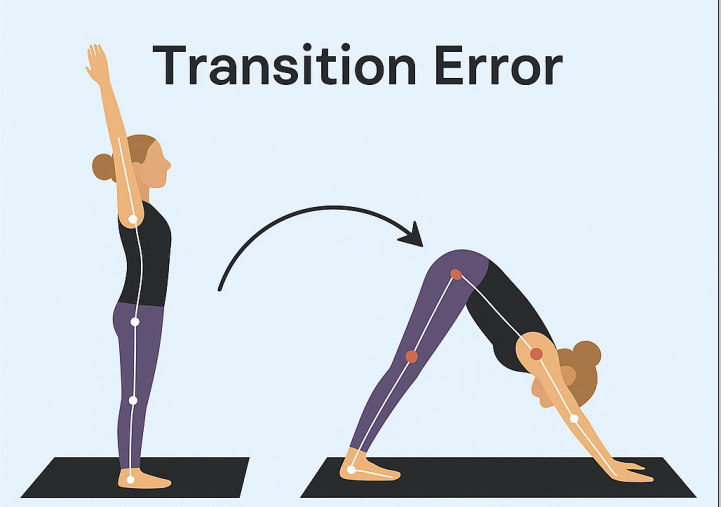}
        \caption{The user's (aggregated) previous performance of pose transition is used to identify if the transition from one pose to the other is smoother or jerky. Based on that, the user can be recommended to follow smoother movements.  }
        \label{fig:transitionerror}
    \end{subfigure}
    \caption{Representative example of error analysis and user feedback for a correct yoga pose \cite{yogahelp-hari}}
    \label{fig:pose-correction}
\end{figure*}
The key errors in Yoga need to be analyzed for better yoga practices :
\begin{itemize}
    \item Pose Error: As SN is a step-by-step process, at each step we stay for a while (1-3 sec in SN), known as a Yoga pose or item. A correct Yoga pose is defined as a deviation from the ideal pose, which is termed as pose error, and can be calculated by the difference between the current pose and the ideal pose.
    \item Transition Error: A transition error is an error made while changing from one pose to another, which requires the analysis of the dynamics of the body. For such analysis, we will use the IMU sensor data. 
\end{itemize} 
These errors can be related to the standard pose data or relative to the user's past performance. The user relative error analysis is done to provide the improvement in the user performance, and over time, to achieve the ideal performance. The user's relative performance is a very effective tool. If a particular exercise is repeated regularly with correct input, the body and muscles become flexible, and overall performance improves day by day. Similar analysis can be performed for the transition error, where our goal is to establish smoother transitions between poses. 

For user recommendation, one can  utilize many techniques to consider advanced problem features, including: 
\begin{itemize}
    \item Knowledge Graph (KG): A KG captures relationships between concepts that can be automatically reasoned with. A KG  can be used in our setting to recommend personalized Yoga items based on the user's health data, such as chronic disease, motor deficiencies, etc. 
    \item ML model: An ML model captures patterns that can be detected from data. An ML model based on live Yoga data analysis can detect and provide real-time feedback to the user. 
\end{itemize}

Furthermore, we want to implement group recommendation methods so that a user is recommended a group of Yoga items subject to their constraints and
preferences. We also want to evaluate its efficacy following best practices in recommendation space  \cite{eval-reco-system}.

\subsection{Sensing }

We want a cost-conscious, accurate, privacy-preserving, secure infrastructure that is easy to use. We also want to optimize sensing by exploring sensor subset selection and optimization - how to choose from the set of sensors for the best results. Finally, we want the management and retrieval of multi-modal data collected to be seamless using convenient REST programming interfaces.

\subsection{Interfaces for Providing Feedback  }
We have to make the system non-invasive, intuitive, and useful. 
The feedback (recommendation) produced by the system itself can be given in a number of ways, such as vibration of the custom device and/or an error bar on an image/ video.  
We also need to evaluate recommendations with human users to measure their effectiveness.

%% file: content/conclusion.tex
\section{Discussion and Conclusion}

In this vision paper, we described the challenge of the problem of Yoga personalization. We identified the challenges, described a preliminary approach, and used it to articulate an outlook for more robust solutions using existing and novel techniques from multidisciplinary computing disciplines. To the best of our
knowledge, this is the first paper that looks at decision support
issues around Yoga personalization comprehensively from
sensing of poses to recommendation of correction for a complete regimen, and illustrates with case study of Surya Namaskar - a set of 12 choreographed poses.

%% file: content/appendix.tex
\clearpage
\onecolumn

\section{SN Performance Analysis from Collected Data}

DAG representation of each variant of SN is shown below in Figures~\ref{fig:sn-state-transition-krishna}, \ref{fig:sn-state-transition-bihar}, \ref{fig:sn-state-transition-vivekananda}.

\begin{itemize}
    \item \textbf{Variant-1 of SN}
\begin{figure}[htbp]
    \centering
    \includegraphics[width=0.68\textwidth]{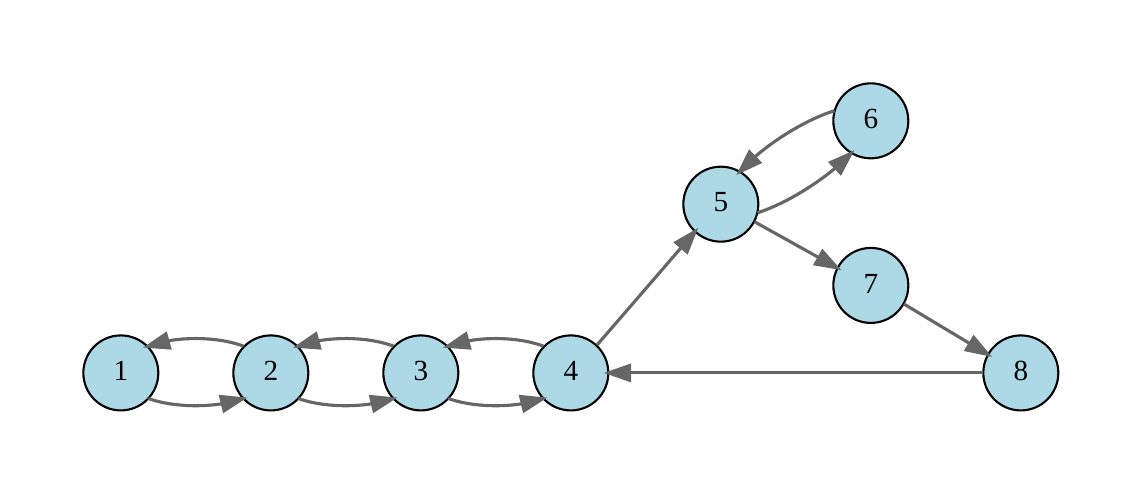}
    \caption{Schematic state-transition representation of Surya Namaskar postures (Krishnamacharya Vinyasa Suryanamaskar).}
    \label{fig:sn-state-transition-krishna}
\end{figure}

\item \textbf{Variant-2 of SN}
\begin{figure}[htbp] 
    \centering
    \includegraphics[width=0.68\textwidth]{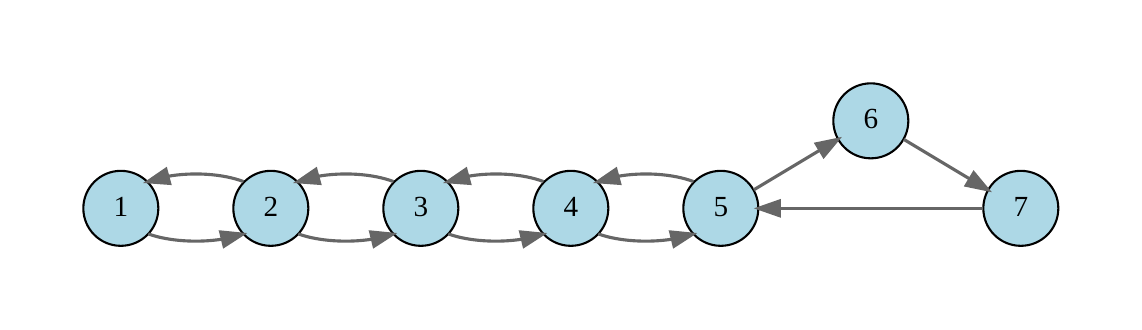} 
    \caption{Schematic state-transition representation of SN postures (Bihar School of Yoga tradition; from \cite{sn-background}.)}
    \label{fig:sn-state-transition-bihar}
\end{figure}

 \item \textbf{Variant-3 of SN}
\begin{figure}[htbp] 
    \centering
    \includegraphics[width=0.68\textwidth]{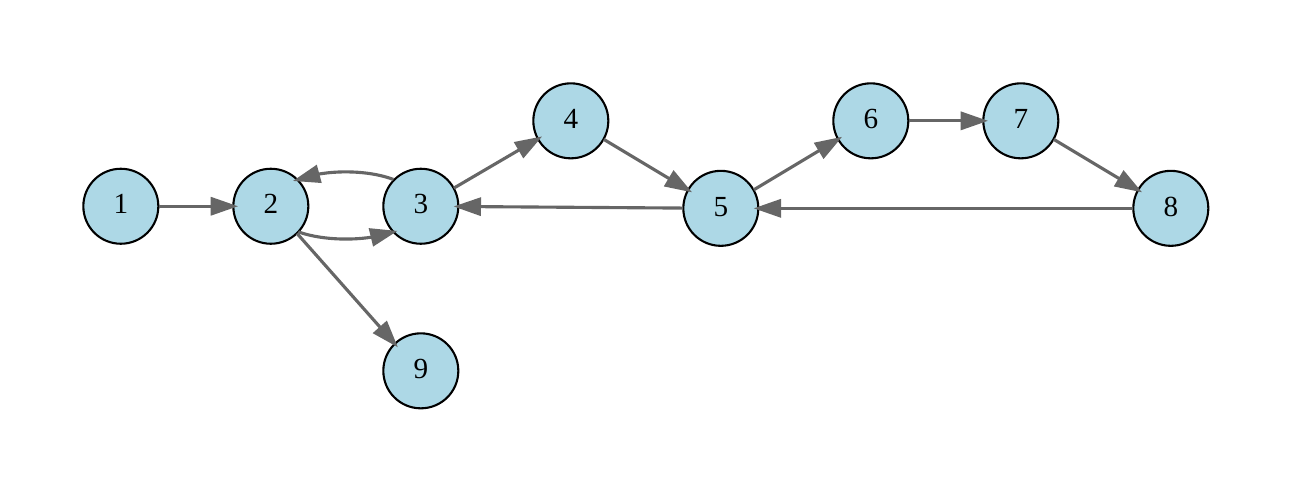} 
    \caption{Schematic state-transition representation of SN postures (Swami Vivekananda Kendra tradition).}
    \label{fig:sn-state-transition-vivekananda}
\end{figure}
\end{itemize}

\section{SN Detailed View}
We show the detailed description of SN across all variants in Table~\ref{tab:full-sn-siva} (see next page).
\begin{table}[htbp]
\footnotesize
\centering
\begin{tabular}{|p{0.5cm}|p{3.3cm}|p{3cm}|p{3.8cm}|p{5.6cm}|}
\hline
\textbf{S.No.} & \textbf{Name} & \textbf{B\={\i}ja Mantra/Full Mantra/Chakra} & \textbf{SN source} & \textbf{Comments}/\textbf{CYP Reference} \\
\hline
1 & Pr\={a}\d{n}\={a}m\={a}sana (prayer pose) & Om Hraam/Om Mitr\={a}ya Namah/Anhata & SN-base, Variant-1, Variant-2, Variant-3(3rd pose) & Called Samasthiti in Variant-1 but done differently. Hence, for variant-1, captured as pose-18. \\
\hline
2 & Hasta utth\={a}n\={a}sana (raised arms pose) & Om Hreem/Om Ravaye Namah /Vishuddhi & SN-base, Variant-1, Variant-2, Variant-3(1st pose) & Called T\={a}\d{d}\={a}sana in Variant-1 but done differently. Hence, for variant-1, captured as pose-17. \\
\hline
3 & P\={a}da hast\={a}sana (hand to foot pose) & Om Hroom/Om S\={u}ry\={a}ya Namah /Swadhisthana & SN-base, Variant-1, Variant-2, Variant-3(2nd pose) & Called Utt\={a}n\={a}sana in Variant-1 but done differently. Hence, for variant-1, captured as pose-21.  \citet[p.~22]{ayush-yoga-info}\\
\hline
4 & A\'{s}va sanc\={a}lan\={a}sana (equestrian pose) & Om Hraim/Om Bh\={a}nave Namah /Ajna & SN-base, Variant-2, Variant-3(3rd pose) & \\
\hline
5 & Parvat\={a}sana (mountain pose) & Om Hraum/Om Khag\={a}ya Namah /Vishuddhi & SN-base (8th pose), Variant-1 (9th pose), Variant-2, Variant-3(8th pose) & \\
\hline
6 & A\d{s}\d{t}\={a}\.{n}ga namask\={a}ra (salute with eight parts) & Om Hrah/Om Pu\d{s}ne Namah /Manipura & SN-base, Variant-2 & \\
\hline
7 & Bhujang\={a}sana (cobra pose) & Om Hraam/Om Hira\d{n}yagarbh\={a}ya Namah /Swadhisthana & SN-base, Variant-2, Variant-3 & \citet[p.~33]{ayush-yoga-info}\\
\hline
8 & Parvat\={a}sana (mountain pose) / Adho mukha \'{s}v\={a}n\={a}sana & Om Hreem/Om Mar\={\i}caye Namah /Vishuddhi & SN-base (8th pose), Variant-1 (9th pose), Variant-2, Variant-3(8th pose) & \\
\hline
9 & A\'{s}va sanc\={a}lan\={a}sana (equestrian pose) & Om Hroom/Om \={A}dity\={a}ya Namah /Ajna & SN-base, Variant-2, Variant-3(3rd pose) & \\
\hline
10 & P\={a}da hast\={a}sana (hand to foot pose) & Om Hraim/Om Savitre Namah /Swadhisthana & SN-base, Variant-1, Variant-2, Variant-3(2nd pose) & \\
\hline
11 & Hasta utth\={a}n\={a}sana (raised arms pose) & Om Hraum/Om Ark\={a}ya Namah /Vishuddhi & SN-base, Variant-1, Variant-2, Variant-3(1st pose) & \\
\hline
12 & Pr\={a}\d{n}\={a}m\={a}sana (prayer pose) & Om Hrah/Om Bh\={a}skar\={a}ya Namah /Anahata & SN-base, Variant-1, Variant-2, Variant-3(3rd pose) & \\
\hline
13 & Caturanga da\d{n}\d{d}\={a}sana & & SN-base (5th pose), Variant-1 (5th or 7th pose), Variant-3(4th pose) & Called Phalak\={a}sana in SN-base \& Variant-3 but done differently. Hence, for SN base, captured as pose-5, variant-3, captured as pose-4. \\
\hline
14 & Utk\={a}\d{t}\={a}sana & & Variant-1 (4th pose) & \\
\hline
15 & Da\d{n}\d{d}a samarpa\d{n}a & & Variant-1 (6th pose) & \\
\hline
16 & \={U}rdhva mukha \'{s}v\={a}n\={a}sana & & Variant-1 (8th pose) & \\
\hline
17 & T\={a}\d{d}\={a}sana & & Variant-1 (12th pose) & Called Hasta utth\={a}n\={a}sana in Variant-1 but done differently. Hence, for variant-1, captured as pose-2.  \citet[p.~20]{ayush-yoga-info} \\
\hline
18 & Samasthiti & & Variant-1 (13th pose) & Called Pr\={a}\d{n}\={a}m\={a}sana in Variant-1 but done differently. Hence, for variant-1, captured as pose-1. \\
\hline
19 & \'{S}a\'{s}\={a}\.{n}k\={a}sana & & Variant-3(5th pose) & \citet[p.~29]{ayush-yoga-info}\\
\hline
20 & \d{S}a\d{s}\d{t}\={a}\.{n}ga namask\={a}ra & & Variant-3(6th pose) & \\
\hline
21 & Utt\={a}n\={a}sana & & Variant-1 (3rd pose or 11th pose) & Called P\={a}da hast\={a}sana in Variant-1 but done differently. Hence, for variant-1, captured as pose-3. \\
\hline
22 & Phalak\={a}sana & & SN-base (5th pose), Variant-3(4th pose) & Called Caturanga da\d{n}\d{d}\={a}sana in SN-base \& Variant-3 but done differently. Hence, for SN base, captured as pose-5, variant-3, captured as pose-4. \\
\hline
\end{tabular}
\caption{Detailed Surya Namaskar pose information. Each pose is traditionally accompanied by a mantra—either a B\={\i}ja (seed) mantra or a full S\={u}rya mantra. The associated chakra indicates the energy center activated by the pose.}
\label{tab:full-sn-siva}
\end{table}

%% file: main.bbl
\begin{thebibliography}{20}
\providecommand{\natexlab}[1]{#1}

\bibitem[{Bakker, Donath, and Rein(2020)}]{exercise-balancing-seniors}
Bakker, J.; Donath, L.; and Rein, R. 2020.
\newblock Balance training monitoring and individual response during unstable vs. stable balance Exergaming in elderly adults: Findings from a randomized controlled trial.
\newblock \emph{Experimental Gerontology}, 139: 111037.

\bibitem[{Basavaraddi(2025)}]{ayush-yoga-info}
Basavaraddi, I.~V. 2025.
\newblock Common Yoga Protocol.
\newblock In \emph{Ministry of Ayush, India}.

\bibitem[{Cramer et~al.(2015)Cramer, Ward, Saper, Fishbein, Dobos, and Lauche}]{yogasafe}
Cramer, H.; Ward, L.; Saper, R.; Fishbein, D.; Dobos, G.; and Lauche, R. 2015.
\newblock The safety of yoga: a systematic review and meta-analysis of randomized controlled trials.
\newblock \emph{American journal of epidemiology}, 182(4): 281--293.

\bibitem[{de~Beukelaar and Dante(2023)}]{exercise-monitoring-survey}
de~Beukelaar, T.~T.; and Dante, M. 2023.
\newblock Monitoring Resistance Training in Real Time with Wearable Technology: Current Applications and Future Directions.
\newblock In \emph{Bioengineering (Basel, Switzerland) vol. 10,9 1085}.

\bibitem[{Garfinkel and Schumacher(2000)}]{GARFINKEL2000125}
Garfinkel, M.; and Schumacher, H.~R. 2000.
\newblock YOGA.
\newblock \emph{Rheumatic Disease Clinics of North America}, 26(1): 125--132.

\bibitem[{Griffith(2013)}]{griffith2013rig}
Griffith, R.~T. 2013.
\newblock \emph{The Rig Veda}, volume~1.
\newblock Library of Alexandria.

\bibitem[{Gupta and Gupta(2021)}]{yogahelp-hari}
Gupta, A.; and Gupta, H.~P. 2021.
\newblock YogaHelp: Leveraging Motion Sensors for Learning Correct Execution of Yoga With Feedback.
\newblock \emph{IEEE Transactions on AI}, 2(4): 362--371.

\bibitem[{Helms et~al.(2020)Helms, Kwan, Sousa, Cronin, Storey, and Zourdos}]{exercise-monitoring-survey-2}
Helms, E.; Kwan, K.; Sousa, C.; Cronin, J.; Storey, A.; and Zourdos, M. 2020.
\newblock Methods for Regulating and Monitoring Resistance Training.
\newblock In \emph{Journal of Human Kinetics, 2020;74:23-42. Published 2020 Aug 31. doi:10.2478/hukin-2020-0011}.

\bibitem[{Irani(2020)}]{yogasafe1}
Irani, G.~S. 2020.
\newblock Injuries/harms resulting from incorrect adjustments/alignments performed by yoga asana practitioners.
\newblock \emph{Archives of Pharmacy Practice}, 11(3-2020): 38--47.

\bibitem[{Isinkaye, Folajimi, and Ojokoh(2015)}]{isinkaye2015recommendation}
Isinkaye, F.~O.; Folajimi, Y.~O.; and Ojokoh, B.~A. 2015.
\newblock Recommendation systems: Principles, methods and evaluation.
\newblock \emph{Egyptian informatics journal}, 16(3): 261--273.

\bibitem[{Li et~al.(2023)Li, Liu, Satapathy, Wang, and Cambria}]{li2023-recentdevelopments-recommendersystems}
Li, Y.; Liu, K.; Satapathy, R.; Wang, S.; and Cambria, E. 2023.
\newblock Recent Developments in Recommender Systems: A Survey.
\newblock arXiv:2306.12680.

\bibitem[{Makaussov et~al.(2020)Makaussov, Krassavin, Zhabinets, and Fazli}]{9283231}
Makaussov, O.; Krassavin, M.; Zhabinets, M.; and Fazli, S. 2020.
\newblock A Low-Cost, IMU-Based Real-Time On Device Gesture Recognition Glove.
\newblock In \emph{2020 IEEE International Conference on Systems, Man, and Cybernetics (SMC)}, 3346--3351.

\bibitem[{Manjunath(2024)}]{yogasafe2}
Manjunath, N.~K. 2024.
\newblock Safety and Prevention of Injuries in Yoga.
\newblock \emph{International Journal of Yoga}, 17(2).

\bibitem[{Nagpal et~al.(2024)Nagpal, Valluru, Lakkaraju, Gupta, Abdulrahman, Davison, and Srivastava}]{mealreco-nagpal2024novelapproachbalanceconvenience}
Nagpal, V.; Valluru, S.~L.; Lakkaraju, K.; Gupta, N.; Abdulrahman, Z.; Davison, A.; and Srivastava, B. 2024.
\newblock A Novel Approach to Balance Convenience and Nutrition in Meals With Long-Term Group Recommendations and Reasoning on Multimodal Recipes and its Implementation in BEACON.
\newblock arXiv:2412.17910.

\bibitem[{TI Sensor Tag()}]{tisensor}
TI Sensor Tag. 2025.
\newblock SimpleLink Bluetooth low energy/Multi-standard SensorTag.
\newblock Texas Instruments.

\bibitem[{Trejo and Yuan(2018)}]{yoga-pose-detection}
Trejo, E.~W.; and Yuan, P. 2018.
\newblock Recognition of Yoga poses through an interactive system with Kinect based on confidence value.
\newblock In \emph{2018 3rd International Conference on Advanced Robotics and Mechatronics (ICARM)}, 606--611.

\bibitem[{UN resolution()}]{internationalyogaday}
UN resolution. 2015.
\newblock Resolution adopted by the General Assembly on 11 December 2014.
\newblock {https://docs.un.org/en/A/RES/69/131}.

\bibitem[{Valluru et~al.(2024)Valluru, Srivastava, Paladi, Yan, and Natarajan}]{ultra-team-group-reco}
Valluru, S.~L.; Srivastava, B.; Paladi, S.~T.; Yan, S.; and Natarajan, S. 2024.
\newblock Promoting research collaboration with open data driven team recommendation in response to call for proposals.
\newblock In \emph{Proceedings of the Thirty-Eighth AAAI Conference on Artificial Intelligence}. AAAI Press.
\newblock ISBN 978-1-57735-887-9.

\bibitem[{Venkatesh and Vandhana(2022)}]{sn-background}
Venkatesh, L.~P.; and Vandhana, S. 2022.
\newblock Insights on Surya namaskar from its origin to application towards health.
\newblock \emph{Journal of Ayurveda and Integrative Medicine}, 13(2): 100530.

\bibitem[{Zangerle and Bauer(2022)}]{eval-reco-system}
Zangerle, E.; and Bauer, C. 2022.
\newblock Evaluating Recommender Systems: Survey and Framework.
\newblock \emph{ACM Comput. Surv.}, 55(8).

\end{thebibliography}
